\documentclass[useAMS,usenatbib,usegraphicx]{mn2e}
\bibpunct[,]{(}{)}{;}{a}{}{}

\usepackage{amssymb}

\title[BCGs and ICL growth]
{The different growth pathways of Brightest Cluster Galaxies and the Intra-Cluster Light}
\author[E.~Contini et al.]
        {E.~Contini,$^{1}$  \thanks{Email: emanuele.contini82@gmail.com} 
         S.K.~Yi,$^{1}$ \thanks{Email: yi@yonsei.ac.kr}
         X.~Kang$^{2}$  \thanks{Email: kangxi@pmo.ac.cn}
         \\ 
         $^1$ Department of Astronomy and Yonsei University Observatory, Yonsei University, Yonsei-ro 50, Seoul 03722, Republic of Korea\\
  	 $^2$ Purple Mountain Observatory, the Partner Group of MPI f\"{u}r Astronomie, 2 West Beijing Road, Nanjing 210008, China \\}
\pagerange{\pageref{firstpage}--\pageref{lastpage}}
\pubyear{2018}

\begin{document}

\maketitle

\label{firstpage}

\begin{abstract} 
We study the growth pathways of Brightest Central Galaxies (BCGs) and Intra-Cluster Light (ICL) by means of a semi-analytic model. We assume that the ICL forms by stellar stripping of satellite 
galaxies and violent processes during mergers, and implement two independent models: (1) one considers both mergers and stellar stripping (named {\small STANDARD} model), and one considers 
only mergers (named {\small MERGERS} model). We find that BCGs and ICL form, grow and overall evolve at different times and with different timescales, but they show a clear co-evolution after redshift 
$z \sim 0.7-0.8$.  Around 90\% of the ICL from stellar stripping is built-up in the innermost 150 Kpc from the halo centre and the dominant contribution comes from disk-like galaxies (B/T$<$0.4) 
through a large number of small/intermediate stripping events ($M_{strip}/M_{sat}<0.3$).
The fractions of stellar mass in BCGs and in ICL over the total stellar mass within the virial radius of the halo evolve differently with time. At high redshift, the BCG accounts for the bulk of the 
mass, but its contribution gradually decreases with time and stays constant after $z\sim 0.4-0.5$. The ICL, instead, grows very fast and its contribution keeps increasing down to the present 
time. The {\small STANDARD} and the {\small MERGERS} models make very similar predictions in most of the cases, but predict different amounts of ICL associated to other galaxies within the virial 
radius of the group/cluster other than the BCG, at $z=0$. We then suggest that this quantity is a valid observable that can shed light on the relative importance of mergers and stellar stripping 
for the formation of the ICL.
\end{abstract}


\section[]{Introduction} 
\label{sec:intro}
The intracluster light (ICL), first discovered by \cite{zwicky37}, is a diffuse light component in galaxy groups and clusters made of stars that are not bound to any galaxy. Since its discovery, 
the ICL has been observed and studied in more detail in order to understand deeper the processes at play in large structures, but only during the last few decades the scientific communinity has 
enriched the knowledge on its formation and evolution. Now it appears clear that the ICL is, somehow, linked to the formation of the brightest cluster galaxies (BCGs) 
(\citealt{murante07,purcell07,puchwein10,rudick11,contini14}, to quote just a few theoretical works and \citealt{demaio15,burke15,groenewald17,morishita17,montes18,demaio18},
to quote the latest observational works), albeit it is still controversial if BCGs and ICL evolve in parallel or the mechanisms responsible of their formation are either different, or 
the same but important at different times. 


BCGs are among the most massive and luminous galaxies known, typically residing in the centre of massive dark matter haloes. Their formation and evolution have been shown to be 
rather different from the less peculiar population of satellite galaxies (\citealt{delucia07}). These massive galaxies form in a hierarchical fashion by accretion and mergers of/with 
satellite galaxies orbiting around them (\citealt{ostriker77,richstone83}), and, according to the most recent theoretical studies, they experience different phases of growth, 
that is, star formation at early times followed by a rapid growth in mass through mergers and multiple accretion of smaller galaxies at later times (e.g., \citealt{delucia07,laporte13,lee17}).
Observational evidence, however, have shown that the rate at which BCGs acquire mass via mergers is still not clear, and their growth in the last 7-8 Gyr is still debated 
among observational studies. In fact, observational works find different growth factor in mass, ranging from no growth (such as \citealt{oliva-altamirano14}), around 50\% (e.g.,
\citealt{lin13}), to almost a factor two (\citealt{lidman12}). 

Many authors (e.g., \citealt{collins09,stott11,burke15,groenewald17}) showed that the observed stellar mass evolution of BCGs disagrees with predictions of models (e.g., \citealt{delucia07}),
and some of them (e.g., \citealt{burke15,groenewald17}) suggested that the tension can be significantly alleviated if a conspicuous fraction, typically around 50\%, of the mass of the 
merging galaxies with the BCGs contributes to build up the ICL. For instance, \cite{groenewald17}, who use the predictions of our model for the ICL formation (\citealt{contini14}, hereafter C14)
and assume a 50\% mass transfer to the BCG during mergers, find that major mergers provide around 30\% of the stellar mass of present-day BCGs since redshift $z\sim 0.45$, and account 
for a sufficient amount of ICL at the same time. However, this argument appears to be in contrast with other recent works (\citealt{montes14,demaio15,demaio18,montes18,morishita17}), which
find that mergers alone cannot explain the observed properties of the ICL, such as color and metallicity.
Very recently, \cite{demaio18} investigate the mechanisms of the ICL formation for 23 galaxy groups and clusters in the redshift range $0.29<z<0.89$, and focus on the color gradients 
of the BCG+ICL, which gets bluer with increasing radius. Their result, together with the fact that the number of mergers required to produce the observed luminosity of the BCG+ICL is 
an order of magnitude larger than expected after $z\sim 1$ (\citealt{lidman13}), indicate that mergers associated with the BCG growth are not the dominant channel for the ICL 
formation.

Violent relaxation during mergers has not been the only mechanism invoked for the formation of the ICL. From the theoretical side, disruption of dwarf galaxies has been another mechanism 
proposed by several authors (\citealt{purcell07,murante07,conroy07}), as well as tidal stripping of intermediate/massive galaxies (\citealt{rudick09,martel12,contini14}), pre-processing/accretion 
from smaller objects (\citealt{rudick06,sommer-larsen06},C14), and \emph{in situ} star formation (\citealt{puchwein10}). Disruption of low-mass galaxies has been ruled out as the dominant 
factor by several works, both theoretical and observational (e.g., \citealt{martel12,montes14,demaio15},C14) since it contributes in terms of number of galaxies involved, but marginally in terms of 
stellar mass that contributes to the ICL. Pre-processing or accretion from smaller objects is a natural consequence of the hierarchical structure formation. Galaxies falling into a larger 
halo bring their own ICL which, because of tidal interactions with the potential well of the new halo, can be easily stripped and became part of the diffuse light associated with the BCG.
Then, the total ICL in the halo increases also through this process. In C14, we show that this mechanism is important for high-mass BCGs, contributing up to 30\% 
of the total ICL for $\log M_{BCG}^* \sim 12$, and less important ($\sim$10\%) for $\log M_{BCG}^* \sim 11$. \emph{In situ} star formation has been proposed by \cite{puchwein10} as a 
possible source of ICL, but there are little, if no evidence, that this channel can be important for the ICL growth (e.g., \citealt{melnick12} find that only 1\% of the ICL can be attributed 
to this channel). 

The scientific community has reached the general consensus that tidal stripping and violent relaxation during mergers are the most important processes for the ICL formation and evolution. 
The ICL assembles the bulk of its stellar mass after $z\sim 1$ (\citealt{murante07},C14) and, as shown by \cite{murante07}, does not seem to have a preferred redshift of formation but rather 
a cumulative process with no favourite timescale. In their simulation, the ICL assembles most of its mass via mergers with the BCG or other massive galaxies, while the contribution 
given by stellar stripping appears to be marginal. According to their results, the formation of the ICL and the build-up of the BCG are strictly connected and parallel.

However, what process between tidal stripping and violent relaxation during mergers plays the most relevant role is not yet confirmed. In C14, we show that stellar 
stripping is responsible for most of the stellar mass in the ICL, while mergers have only a marginal role. Most of the ICL comes from stripping of intermediate/massive galaxies, around 70\% 
from galaxies with mass $\log M_* >10.5$. This is in good agreement with recent observations focused on the ICL metallicity (e.g., \citealt{demaio15,montes18,harris17}), that are similar to those 
found in the outskirts of the Milky Way (\citealt{cheng12}), and colors (e.g., \citealt{zibetti05,montes14,iodice17}). As explained in C14, more massive galaxies are more metal rich and, because 
of dynamical friction, they get closer to the centre of the cluster in a shorter timescale (\citealt{presotto14,roberts15,contini15}).

In this work, we make use of the Tidal Radius model described in C14 and another model which considers only violent relaxation during merger as the channel for the formation of the ICL, to 
address the following questions:
\begin{itemize}
 \item What is the relative contribution of mergers and stellar stripping to the BCGs and ICL growth?
 \item Do the growth pathways of BCGs and ICL go in parallel?
 \item Does the ICL formation have a preferred timescale or it is a linear process with time?
\end{itemize}

In Section \ref{sec:methods}, we briefly summarise our modelling, from the main features of the set of simulations used, to the detail of the two prescriptions used for the formation of the ICL. 
In Section \ref{sec:results}, we show the results of our analysis, which will be discussed in detail in Section \ref{sec:discussion}. In Section \ref{sec:conclusion}, we give our conclusions.
Throughout this paper we use the standard cosmology summarized in Section \ref{sec:methods}. The stellar masses are given in units of $M_{\odot}$ (unless otherwise stated), and we assume a 
\cite{chabrier03} IMF.

\section[]{Methods}  
\label{sec:methods}

As in C14, in this study we use a semi-analytic approach. We couple the same semi-analytic model (\citealt{delucia07}), updated with prescriptions for stellar stripping and violent relaxation 
during mergers, with a set of high-resolution N-body simulations. We briefly summarise here the main features of the approach/modelling, and refer the reader to C14 for more details.

\subsection[]{Simulations and Mergers Trees}
\label{sec:simulations}
We use 27 high-resolution numerical simulations of regions around galaxy clusters (the same set used in C14), and the simulation data have been stored at 93 output times, between $z=60$ and $z=0$.

For further details on the simulations and their post-processing, we refer the reader to \citet{contini12} or C14. Subhaloes have been populated with galaxies by means of the above quoted semi-analytic model
(SAM), which uses the information stored in the merger trees to treat the formation and evolution of the galaxy population according to several physically motivated prescriptions that consider most of the baryonic 
physics at play, including cooling of gas, star formation and disk instabilities, mergers, SN and AGN feedback. At any redshift, the SAM populates any new main halo with an amount of hot gas that depends 
on the cosmic baryon fraction. The hot gas, or part of it, can eventually cool afterwords, via cooling processes. The amount of hot gas converted to cold gas depends mainly on the properties of the host halo. Once some cold gas 
is present, star formation can occour leading to a galaxy with stellar mass, cold and hot gas mass. The evolution from that moment on depends on the particular merging history of the host halo. As long as the 
host halo is a main halo, the galaxy is considered a central. If the host becomes part of another halo (i.e. it becomes a subhalo), the galaxy is considered a satellite. Centrals and satellites are treated differently 
in many aspects for reasons that go beyond the scope of this paper. For further details, we refer the reader to \cite{delucia04} and \cite{delucia07}. 

\subsection[]{Prescriptions for the ICL Formation}
\label{sec:models}

In order to address the purposes of this study, we use two prescriptions for modelling the formation of the ICL. For the sake of simplicity, we refer to them as {\small STANDARD} and {\small MERGERS} 
models. The {\small STANDARD} model is equivalent to the Tidal Radius + Merg. model described in C14, which considers both stellar stripping and violent relaxation during mergers. The {\small MERGERS}
model is just a revision of the prescription for the formation of the ICL during mergers presented in C14, adjusted to the most recent assumptions about the percentage of stellar mass that becomes unbound 
during a merger event. In the following we give the detail of each prescription and motivation of the assumptions used.

{\bf STANDARD}: this model is formally identical to the Tidal Radius + Merg. model adopted in C14. Given the importance of a full comprehension of the prescription for the scope 
of this work, here we describe in detail the major features of the modelling. The {\small STANDARD} model takes into account stellar stripping due to tidal interactions between satellite galaxies 
and the potential well of the group/cluster within which they reside, and violent relaxation during galaxy mergers. We model stellar stripping by allowing each satellite galaxy to lose mass in a continuous
fashion, before merging or being totally destroyed if the tidal field is strong enough. We assume that a spherically symmetric isothermal profile can approximate the stellar density distribution of each 
satellite. Then, the \emph{tidal radius} can be estimated by means of the equation (see \citealt{binney}):

\begin{equation}\label{eqn:tid_rad}
  R_{t} = \left(\frac{M_{sat}}{3 \cdot M_{DM,halo}}\right)^{1/3} \cdot D \, ,
\end{equation} 
where $M_{sat}$ is the satellite mass (stellar mass + cold gas mass), $M_{DM,halo}$ is the dark matter mass of the parent halo, and $D$ the satellite distance from the halo centre. 

The model considers a galaxy as a two-component system with a spheroidal component (the bulge), and a disk component \footnote{An isothermal profile is implicitly assumed by the SAM to derive the tidal radius 
via Equation \ref{eqn:tid_rad}. However, for a more realistic implementation of stellar stripping, a galaxy is considered to be a two-component system when stellar stripping occours.}. The satellite galaxy which 
experiences some kind of tidal interaction, can lose mass in two ways: 
\begin{itemize}
 \item [(a)] if $R_t$ is smaller than the bulge radius, the satellite is assumed to be completely disrupted and its stellar and cold gas mass to be added to the ICL and hot component of the central galaxy, 
           respectively;
 \item [(b)] if $R_t$ is larger than the bulge radius but smaller than the disk radius, the model moves the stellar mass in the shell $R_t -R_{sat}$ to the ICL component of the central galaxy. For 
           consistency, also a proportional fraction of the cold gas in the satellite galaxy is moved to the hot component of the central galaxy.
\end{itemize}
We assume an exponential profile for the disk, and $R_{sat} = 10 \cdot R_{sl}$, where $R_{sl}$ is the disk scale length ($R_{sat}$ thus contains 99.99 per cent of the disk stellar mass). After a stripping episode
(b), the disk scale length is updated to one tenth of the $R_t$. For more detail about the prescription for the bulge and disk sizes, we refer the reader to Section 3.5 of C14.

Our semi-analytic model treats two different kinds of satellite galaxies: satellites still associated with a parent subhalo (a.k.a. type1 in our terminology), and satellites which lost their own subhalo, 
because either it has been stripped or the number of particles associated to it went below the resolution of the simulation (a.k.a. type2 or orphans in our terminology). For the latter, Eq. \ref{eqn:tid_rad} 
above is applied directly with no other filter, but for type1 satellites, the model first requires that the following condition is met:

\begin{equation}\label{eqn:eq_radii}
 R^{DM}_{half} < R^{Disk}_{half} \, ,
\end{equation}
where $R^{DM}_{half}$ is the half-mass radius of the parent subhalo, and $R^{Disk}_{half}$ the half-mass radius of the galaxy's disk, that is $1.68\cdot R_{sl}$ for an exponential profile. Anytime a type1 satellite 
is affected by stellar stripping, the associated ICL component is added to that of the corresponding central galaxy. 

The contribution to the formation of the ICL due to violent relaxation processes that take place during mergers is modelled in a very simple fashion. We assume that, when two galaxies merge, a fraction $f_m =0.2$ of the
satellite stellar mass becomes unbound and is added to the ICL of the corresponding central galaxy. In C14, that fraction $f_m$ has been chosen in order to reproduce, approximately, the results of the numerical 
simulations by \cite{villalobos12} \footnote {In reality, $f_m$ might be a function of sevaral properties of the merging galaxies (e.g., stellar mass ratio between satellite and central, orbital 
circularity, etc. etc.), or properties of the haloes involved (such as mass ratio between the parent haloes of satellite and central). In a forthcoming paper, we aim to use a control set of cluster 
simulations to infer an empirical formula capable to link the fraction of mass lost to the above mentioned quantities.}.

{\bf MERGERS}: this model considers only the violent relaxation processes during galaxy mergers as a channel for the formation of the ICL, i.e. \emph{no stellar stripping is implemented}. The prescription is as simple 
as the ``merger channel" in the previous model, with the difference that, in this case, $f_m$ is set to 0.5. Then, the assumption is that, during a merger, the stellar mass of the satellite galaxy is split 
equally between the central galaxy and the ICL associated to it. We have chosen such a fraction for two reasons: (a) that is the most common assumption maden in recent observational studies to justify the growth of the 
ICL from intermediate redshifts to the present day (see, e.g., \citealt{burke15,groenewald17}), and (b) in C14 we have verified that assuming a larger fraction of the satellite stellar mass that gets unbound does not 
affect further (with respect to the case of $f_m =0.5$) the ICL fraction because the effect of having more stellar mass unbound is balanced by the fact that merging galaxies become significantly less massive (see also 
\citealt{moster18}).

However, it must be noted that there is at least another possible way to define the merger channel, and it relies on the very last moments spent by satellites before merging with the central. In fact, the merger 
channel in SAMs is usually linked to the very last moment, i.e. when the two galaxies are considered to be one system, while in simulations the merger is a more continuous process in time. Then, another way to define 
the merger channel would be by considering what happens to the satellite stellar mass in the very vicinity of the central, before actually merging with it. These last moments are included by our model in the stripping 
channel, rather than in the merger channel. We will fully discuss the implications of it in Section \ref{sec:discussion}.

\section{Results}
\label{sec:results}

In this section, we aim to address the questions itemized in Section \ref{sec:intro} by looking at the predictions of our two models and the intrinsic differences among them. In particular, we intend to achieve our
goals by analyzing the contributions to the ICL and BCGs from mergers and stellar stripping, and their growths from high-redshift to the present time. Our sample of haloes is the one studied in C14, where the 
halo mass ranges between $\log M_{200} \sim 13.0$ and $\log M_{200} \sim 15.0$, with a total number of 341 haloes (and so BCGs).

In C14 (Fig. 2) we find that the fraction of ICL as a function of halo mass predicted by the ``Tidal Radius + Merg." model (current {\small STANDARD} model) ranges between 20\%-30\% and, considering the scatter,
with no halo mass dependence. Our {\small MERGERS} model predicts the same trend, but slightly lower fraction, between 15\% and 20\% (result not shown). As argued in C14, the scatters in the model and in the 
observed fractions make these results both in agreement with observations, ruling-out the chance to choose a favourite model. One possibility to address this issue might be given by the amount of ICL that is 
associated with the BCG \footnote{In C14 we assumed centrals and type1 satellites to have their own ICL component. For central galaxies, their associated ICL can come from direct stripping or mergers with 
satellites, or be accreted from galaxies falling into the cluster. In the latter case, the ICL of these galaxies is instantaneously stripped and added to the ICL component of the central galaxy if the satellite 
is a type2 (orphan), while the same happens at the first episode of stripping if the satellite is a type1.}(i.e. only the ICL around the central object of the halo) compared with the total amount of ICL in the 
halo (that is, instead, associated with other galaxies), at $z=0$. In fact, in the {\small MERGERS} model, type1 satellites maintain their own ICL because there is no stripping, while in the {\small STANDARD} model 
they can lose their ICL after the first stripping event. For this reason, BCGs are expected to have a larger amount of ICL in the case of the {\small STANDARD} model.

\begin{figure} 
\begin{center}
\includegraphics[scale=.45]{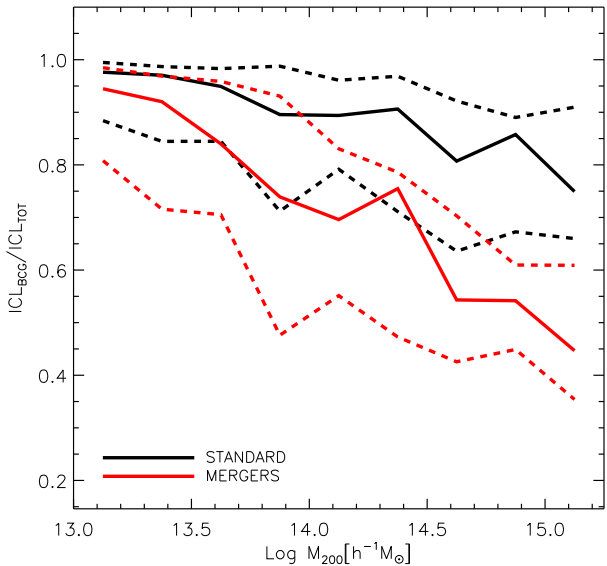} 
\caption{Fraction of the ICL associated with the BCG over the total ICL within $R_{200}$ associated with other galaxies, as 
a function of halo mass. The black lines represent the prediction of the STANDARD model, while the red lines represent the prediction of the MERGERS model. Solid 
lines represent the median of the distribution, and dashed lines represent the 16$^{th}$ and 84$^{th}$ percentiles.}
\label{fig:iclbcg_icltot}
\end{center}
\end{figure}

In Figure \ref{fig:iclbcg_icltot} we plot the fraction of stellar mass in the ICL associated with the BCG, over the total amount of ICL in the group/cluster, 
as a function of the virial mass of the halo. Black lines show the median, 16$^{th}$ and 84$^{th}$ percentiles of the distribution predicted by the {\small STANDARD} model, while the red lines show the same quantities 
predicted by the {\small MERGERS} model. In the mass range of small groups ($<10^{13.5} M_{\odot}/h$), the two models predict similar fractions, but, as the mass of the halo increases, the fraction predicted by 
the {\small MERGERS} model decreases faster than that predicted by the {\small STANDARD} model, until they are perfectly distinguishable on high-mass cluster scale. Hence, specific and deep observations focused on 
cluster scale at the present time might shed some light on this and favour one or another model. Taken note of that, in the following we present the results of our analysis by comparing the predictions of the two models.

\subsection[]{Contribution from Mergers}
\label{sec:merg_contr}

\begin{figure*} 
\begin{center}
\includegraphics[scale=.8]{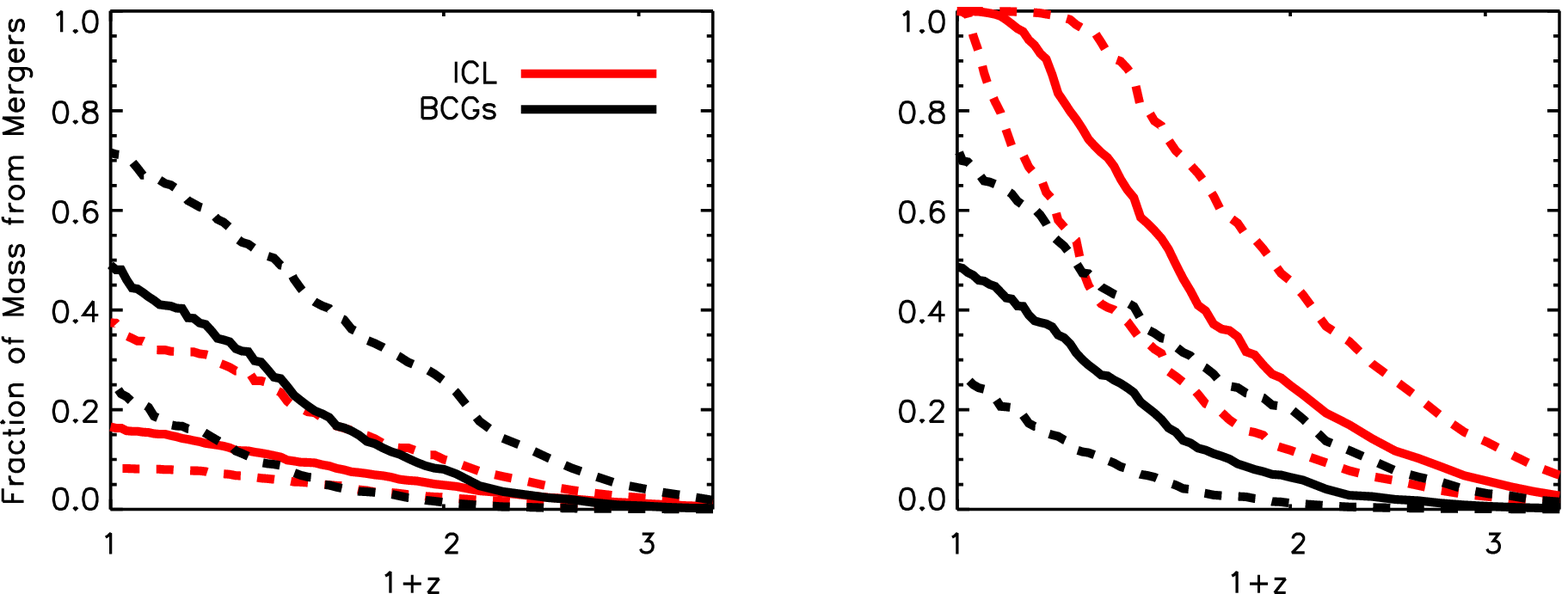} 
\caption{Cumulative fraction of stellar mass gained via mergers by the ICL (red lines) and by the BCGs (black lines) normalised to their mass at $z=0$, as a function of redshift. The left panel refers to 
the STANDARD model, while the right panel refers to the MERGERS model. Solid lines represent the median of the distribution, and dashed lines represent the 16$^{th}$ and 84$^{th}$ percentiles.}
\label{fig:fmerg_all}
\end{center}
\end{figure*}

Figure \ref{fig:fmerg_all} shows the cumulative fraction of stellar mass built-up by BCGs (black lines) and ICL (red lines) through mergers, as a function of redshift. The left panel shows the predictions of the
{\small STANDARD} model, while the right panel shows the predictions of the {\small MERGERS} model. Solid lines represent the median of the distribution, and dashed lines represent the 16$^{th}$ and 84$^{th}$ 
percentiles. In the case of the {\small STANDARD} model, BCGs build-up a large amount of mass down to redshift $z \sim 1$ and, on average, 50\% of their present mass comes from mergers. Moreover, the scatter shown 
by the black dashed line increases with decreasing redshift, and ranges between 25\% and 70\%, at $z=0$. The picture is different for what concerns the ICL. The fraction grows gradually as time passes, and reaches a 
median value of $\sim$ 17\% at $z=0$, and the scatter associated to the ICL ranges from 10\% to almost 40\%. The predictions of the {\small MERGERS} models do not qualitatively and quantitatively change 
the trend shown by the {\small STANDARD} model in the case of BCGs, while radically change that of the ICL. This is obvious since, by construction, in this model the ICL can grow only through mergers.

\begin{figure*} 
\begin{center}
\includegraphics[scale=.8]{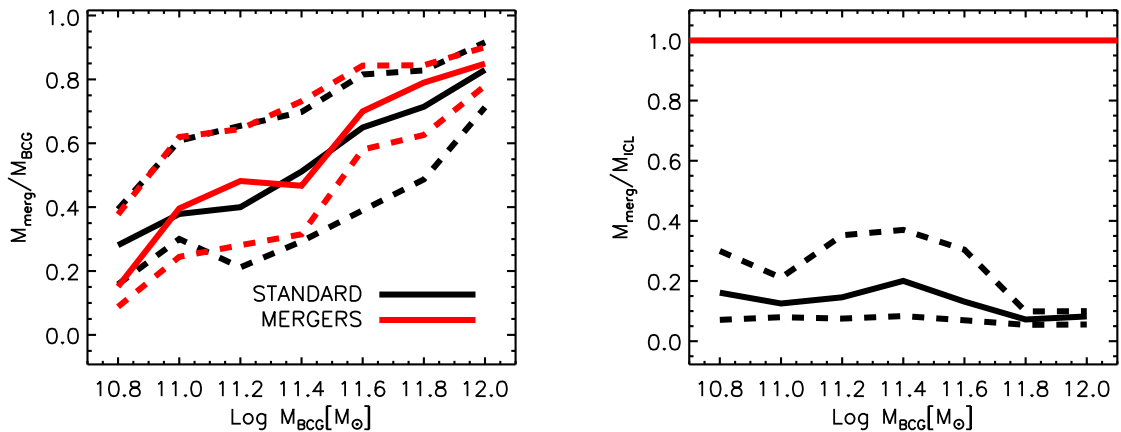} 
\caption{Left panel: fraction of stellar mass gained via mergers by BCGs down to present time as a function of the BCG mass at $z=0$, as predicted by the STANDARD model
(black lines) and by the MERGERS model (red line). Right panel: same as the left panel, but for the ICL. Solid lines represent the median of the distribution, and dashed 
lines represent the 16$^{th}$ and 84$^{th}$ percentiles.}
\label{fig:fmerg_bcg}
\end{center}
\end{figure*}

\begin{figure*} 
\begin{center}
\begin{tabular}{cc}
\includegraphics[scale=.8]{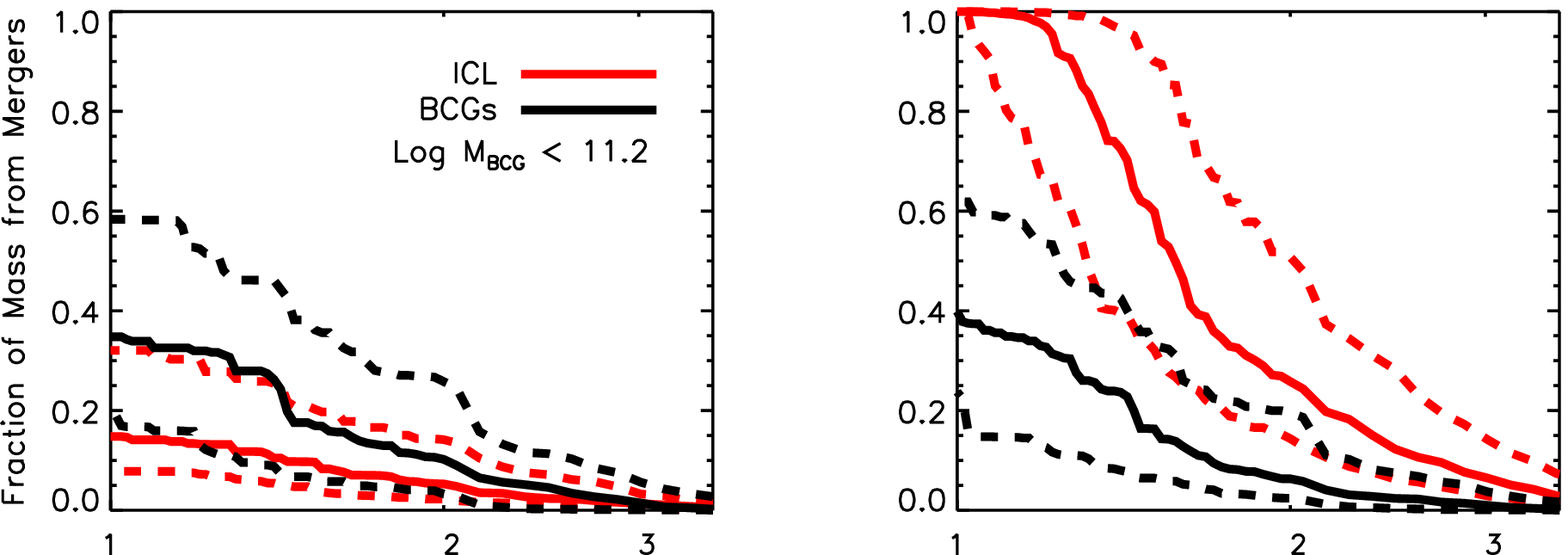} \\
\includegraphics[scale=.8]{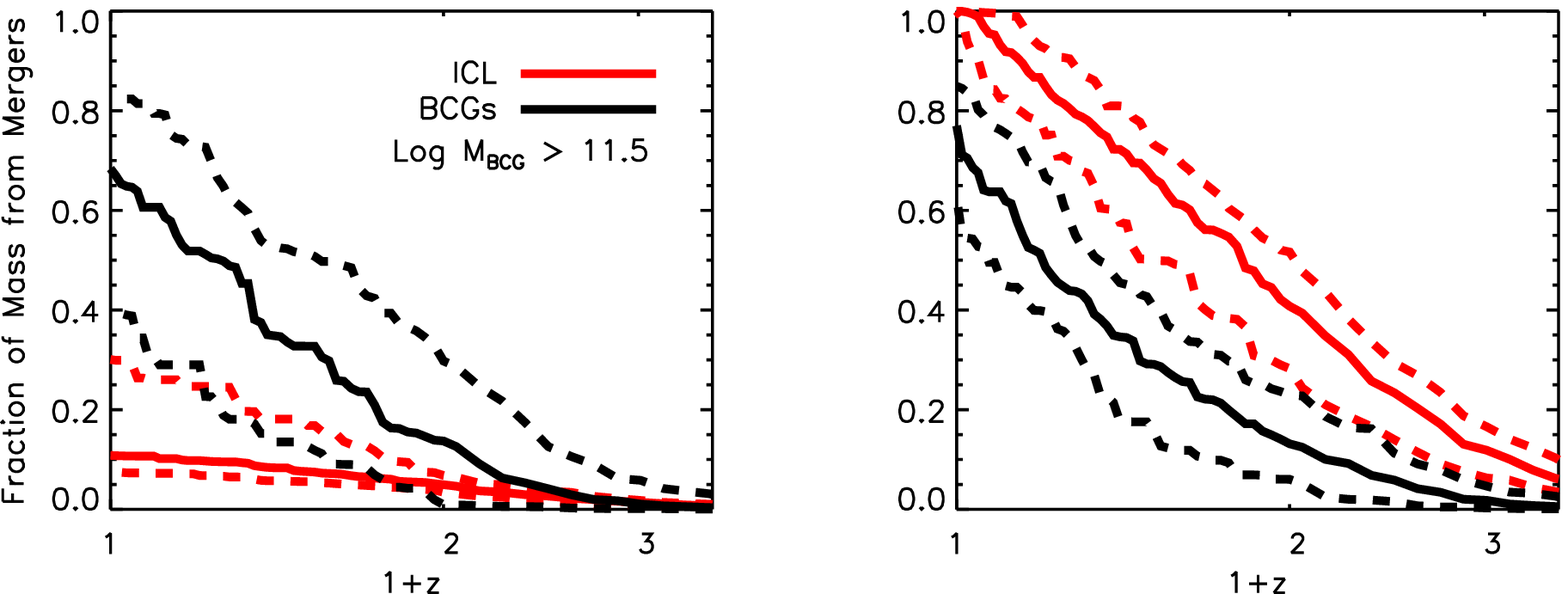} \\ 
\end{tabular}
\caption{Same as Figure \ref{fig:fmerg_all}, but for BCGs with $\log M_* [M_{\odot}] <11.2$ and BCGs with $\log M_* [M_{\odot}] >11.5$.}
\label{fig:fmerg_lowhigh}
\end{center}
\end{figure*}

The scatter seen in Figure \ref{fig:fmerg_all} leaves the possibility that part of it is due to different mass accretion histories of BCGs, that is, there might be a possible dependence on the final BCG stellar mass.
In the left panel of Figure \ref{fig:fmerg_bcg}, we show the ratio between the total stellar mass gained by mergers and the stellar mass at $z=0$ for BCGs, as a function of the latter. The black and red colors refer 
to the {\small STANDARD} and {\small MERGERS} models, respectively. Two aspects of this plot must be noted: this ratio is a clear function of the final BCG mass, and there is no dependence on the modelling. Interestingly,
in around one order of magnitude in stellar mass, from $\log M_{BCG}^* \sim 10.8$ to $\log M_{BCG}^* \sim 12.0$, the ratio increases from $\sim 0.2$ to $\sim 0.8$. More massive BCGs build-up a larger amount of stellar mass via 
mergers than less massive BCGs, consistent with another SAM prediction (\citealt{lee13,lee17}) and the hydro-simulation Illustris (\citealt{rodriguez-gomez16}). This is a consequence of the fact that more massive BCGs 
are supposed to reside in larger haloes and, given their potential wells and the larger number of satellites, this BCGs experience more mergers than the less massive ones. The right panel of Figure \ref{fig:fmerg_bcg} shows 
the same quantity for the ICL, including the case of the {\small MERGERS} model where the ratio is one by definition. The ICL built-up from mergers has no favourite mass scale in the {\small STANDARD} model, given the 
fact that the ratio is almost constant. This is not a new result because the right panel of Figure \ref{fig:fmerg_bcg} is identical to the botton panel of Figure 7 in C14. 

In Figure \ref{fig:fmerg_lowhigh} we address the point discussed above, that is, the dependence of the mass gained via mergers by BCGs on their final mass. For our purpose, from the whole sample of BCGs (341),
we select two subsamples: BCGs with mass lower than $\log M_{BCG}^* \sim 11.2$ (83), and higher than $\log M_{BCG}^* \sim 11.5$ (67). The upper panels of Figure \ref{fig:fmerg_lowhigh} show the same quantities shown in Figure 
\ref{fig:fmerg_all} for the least massive BCGs, and the bottom panels for the most massive BCGs. Colors and linestyles have the same meaning as in Figure \ref{fig:fmerg_all}, for the {\small STANDARD} (left panels) and 
{\small MERGERS} (right panels) models. These plots highlight an interesting aspect. The BCG growth does depend on the final mass of the BCGs, as partially anticipated above. Larger BCGs end up at the present time with 
double the percentage (70\%) of stellar mass gained via mergers than lower mass BCGs (35\%) in the case of the {\small STANDARD} model \footnote{Note that the ranges of the values in Figure \ref{fig:fmerg_all} are different 
from those shown in Figure \ref{fig:fmerg_lowhigh} because the whole sample of BCGs does not coincide with the two subsamples together. Our subsamples have been chosen in such a way to accentuate the intrinsic differences 
(between them) of the quantities analysed.}. The trend is still the same in the case of the {\small MERGERS} model, with slightly 
different median values and scatters. The ICL, on the contrary, does not show particular dependence on the BCG stellar mass in the predictions of the {\small STANDARD} model, while it shows little different evolutions 
in the case of the {\small MERGERS} model. It would be reasonable to expect a larger amount of ICL via mergers from the {\small STANDARD} model for the most massive BCGs, since this is also the case for the BCG stellar mass.
The ICL associated to larger BCGs does build-up more stellar mass from mergers (with respect to the ICL associated to less massive BCGs), but this is counterbalanced by the fact that it also builds up more stellar mass 
through stellar stripping (C14). 

\subsection[]{Contribution from Stellar Stripping}
\label{sec:strip_contr}

\begin{figure*} 
\begin{center}
\includegraphics[scale=.8]{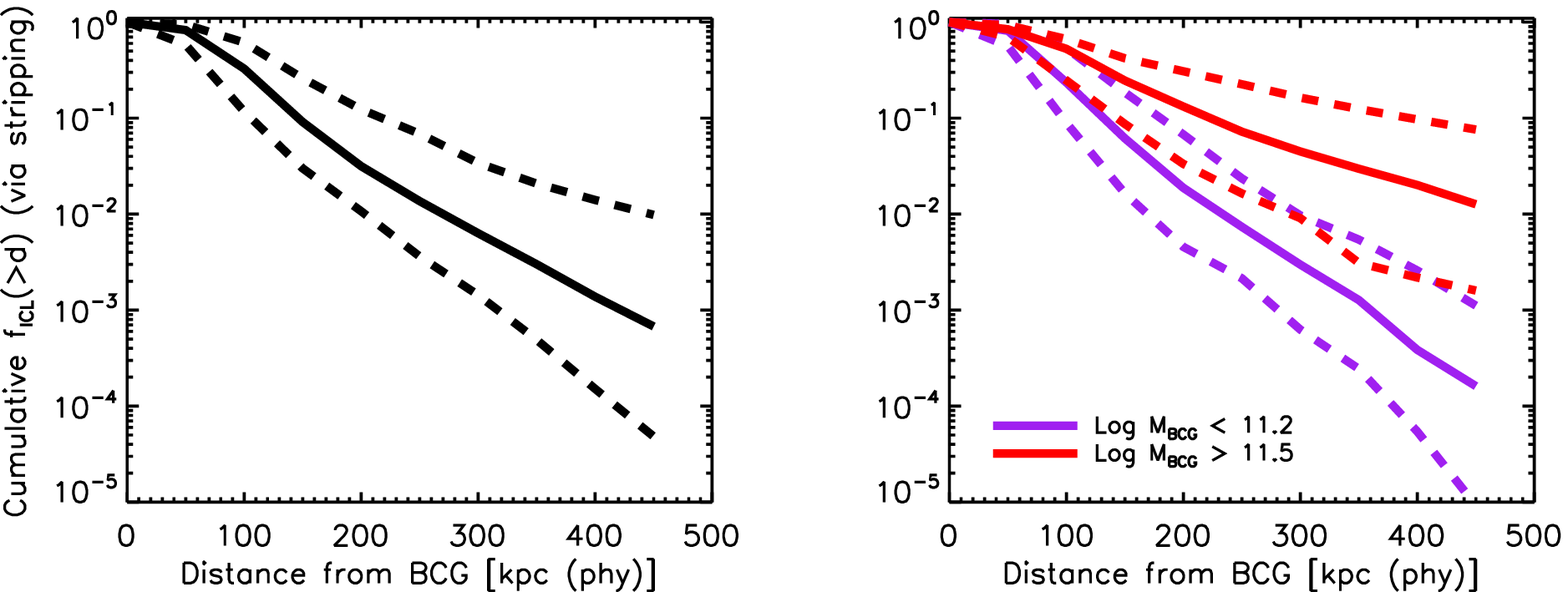} 
\caption{Left panel: cumulative fraction of stellar mass in ICL gained via stripping as a function of distance (from the halo centre) of the satellite galaxies from which the mass 
has been stripped for the whole sample of BCGs. Right panel: same as the left panel, but for BCGs with $\log M_* [M_{\odot}] <11.2$ (purple lines) and BCGs with $\log M_* [M_{\odot}] >11.5$ 
(red lines). Solid lines represent the median of the distribution, and dashed lines represent the 16$^{th}$ and 84$^{th}$ percentiles. The plots show the prediction of the STANDARD 
model, because in the MERGERS model no ICL forms via stellar stripping.}
\label{fig:iclstrip}
\end{center}
\end{figure*}

\begin{figure*} 
\begin{center}
\begin{tabular}{cc}ß
\includegraphics[scale=.4]{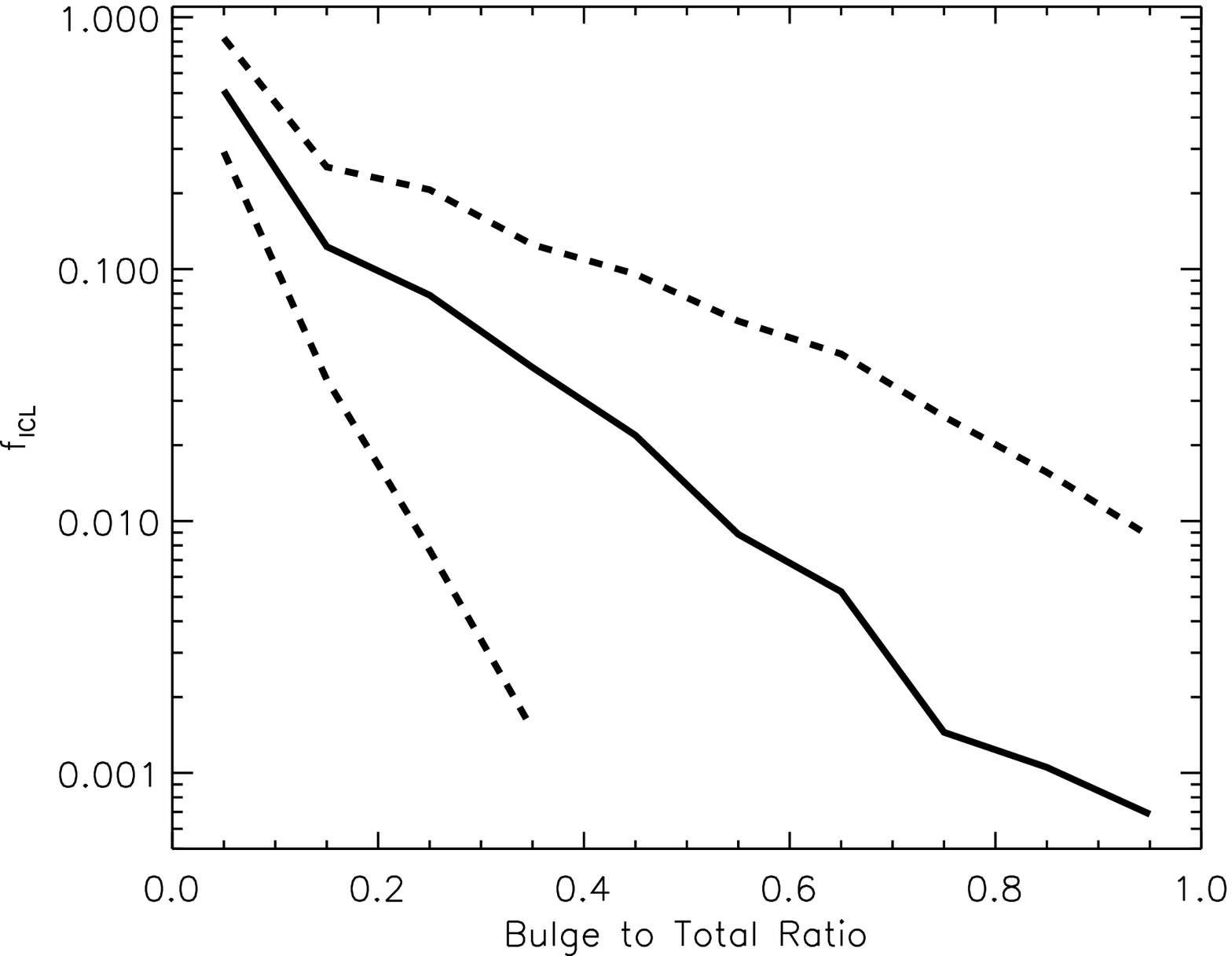} &
\includegraphics[scale=.4]{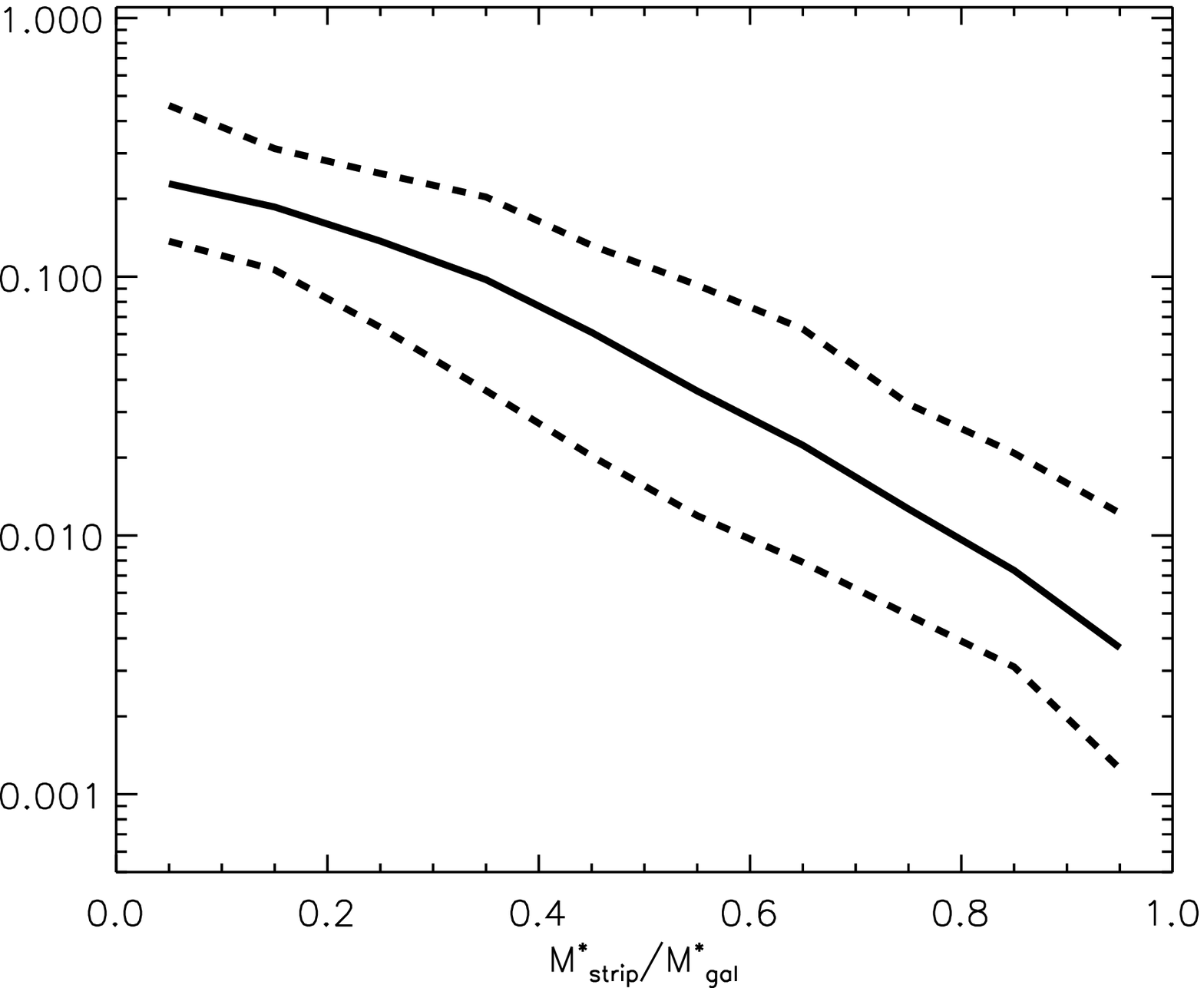} \\ 
\end{tabular}
\caption{Left panel: fraction of ICL mass via stellar stripping as a function of the bulge-to-total mass ratio of satellite galaxies involved in the stripping events, within 100 Kpc from 
the centre of the halo. Right panel: same as the left panel, but as a function of the ratio between the mass stripped and the mass of the satellite galaxy from which the mass has been 
stripped. Solid lines represent the median of the distribution, and dashed lines represent the 16$^{th}$ and 84$^{th}$ percentiles. As in Figure \ref{fig:iclstrip}, the plots 
show only the prediction of the STANDARD model.}
\label{fig:icl_BT}
\end{center}
\end{figure*}

The left panel of Figure \ref{fig:iclstrip} shows the cumulative fraction of ICL stellar mass built-up exclusively from stellar stripping, as a function of the distance (from the halo centre) of the 
satellite galaxies from which the mass has been stripped, and for the whole sample of BCGs. The solid line represents the median of the distribution and the dashed lines indicate the 16$^{th}$ 
and 84$^{th}$ percentiles. The message of this plot is quite evident: on average, only around 10\% of the total ICL mass built-up via this channel comes from stripping events happened farther than
150 Kpc from the halo centre. It means that, by considering the whole sample of BCGs, around 90\% of their ICL from stellar stripping is built-up in the first 150 Kpc, in a region relatively small
around the BCG. However, the total scatter appears large, and, at 150 Kpc, the percentage ranges from 74\% to 97\%. This suggests that this quantity might be a function of the stellar mass 
growth of the BCGs to which the ICL is associated, hence, on the final BCG mass at the present time. This dependence is shown in the right panel of Figure \ref{fig:iclstrip}, for BCGs with 
$\log M_* <11.2$ (purple lines) and BCGs with $\log M_* >11.5$ (red lines). The solid purple line stays below the red solid line at all distances except in the very central 
regions (within $\sim$ 50 Kpc), meaning that, on average, the contribution of stellar stripping events at larger distance than 50 Kpc for the built-up of the ICL is much more effective for larger 
BCGs than for the less massive ones. For example, more massive BCGs build-up 75\% of their ICL (from stellar stripping) in the inner 150 Kpc, while at the same distance, the percentage increases to 94\%
for the less massive BCGs. We believe that this difference is the result of a well known trend, that is, less massive BCGs reside in less massive haloes, which are more centrally concentrated than larger 
haloes (see, e.g.,\citealt{gao11,contini12}). In the central regions, stellar stripping becomes more effective, and even more for less massive haloes (see also C14 for more details).

If we restrict the analysis within the first 100 Kpc from the centre, the average percentages become 67\%,77\% and 47\%, considering the whole sample of BCGs, less massive and more massive BCGs, respectively. 
In C14 we have shown that intermediate/massive satellite galaxies contribute most to the formation of the ICL, and we have explained it with dynamical arguments. Here, we want to focus on the morphology 
of galaxies and on their relative contribution in terms of actual mass liberated during each stripping event in the very central regions, within 100 Kpc. In Figure \ref{fig:icl_BT} we show the fraction 
of ICL (via stellar stripping) as a function of the bulge-to-total mass ratio (left panel), and as a function of the ratio between the actual stellar mass lost and the mass of the satellite subject to 
stripping (right panel). The left panel clearly shows that disk-like galaxies are those which contribute most to the ICL formed via stripping in the innermost region centred around the central object, while
elliptical/spheroidal galaxies (B/T$>0.4$) contribute marginally, no more than 25\%. This is a consequence of our modelling, but supported by physical reasons (see \citealt{chang13}). In fact, two galaxies, 
a disk-like and a bulge-like, with the same stellar mass and positioned at the same distance, have different probability to be stripped. According to Equation \ref{eqn:eq_radii}, the tidal radius for these 
two galaxies would be the same, but the size can be very different being the disk more extended then the bulge. Then, disk-like galaxies have an higher probability to be stripped. However, it must be noted 
that this result has to be taken with caution because semi-analytic models, in general, predict a larger fraction of elliptical galaxies (with B/T$>0.7$, see e.g. \citealt{wilman13}).

The right panel of Figure \ref{fig:icl_BT}, instead, shows that a large number of small/intermediate stripping events (mass fractions below 0.3) account for most of the ICL coming from this channel, and 
the total distruption of the satellite is statistically unlikely. On average, small/intermediate stripping events account for 83\% of the ICL accumulated within 100 Kpc from the centre and, as discussed 
above, they mainly involve disk-like galaxies. Given the fact that 100 Kpc is a sufficient small area around the BCG (in most of the cases the disk of BCGs extends even farther), we can state that this 
stripping events actually involve also galaxies in the process of merging. We will discuss this argument in more detail in Section \ref{sec:discussion}.

\subsection[]{ICL and BCG Growth}
\label{sec:growth}

\begin{figure*} 
\begin{center}
\begin{tabular}{cc}
\includegraphics[scale=.8]{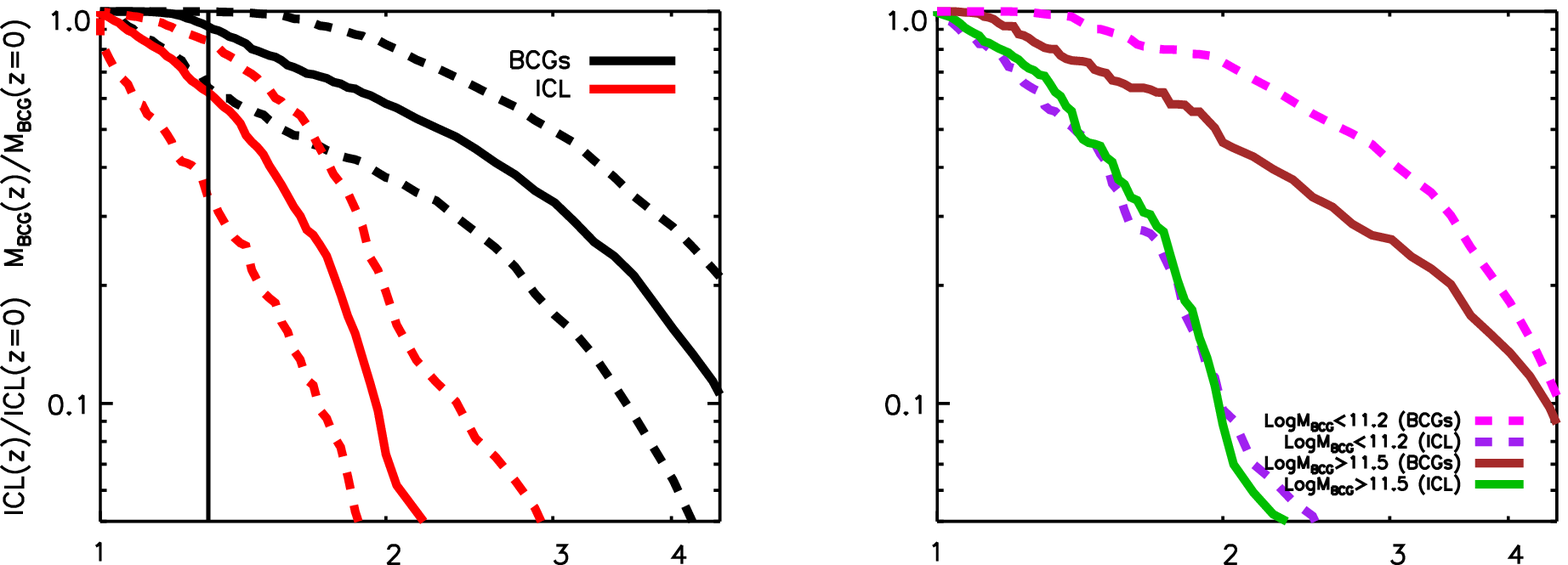} \\
\includegraphics[scale=.8]{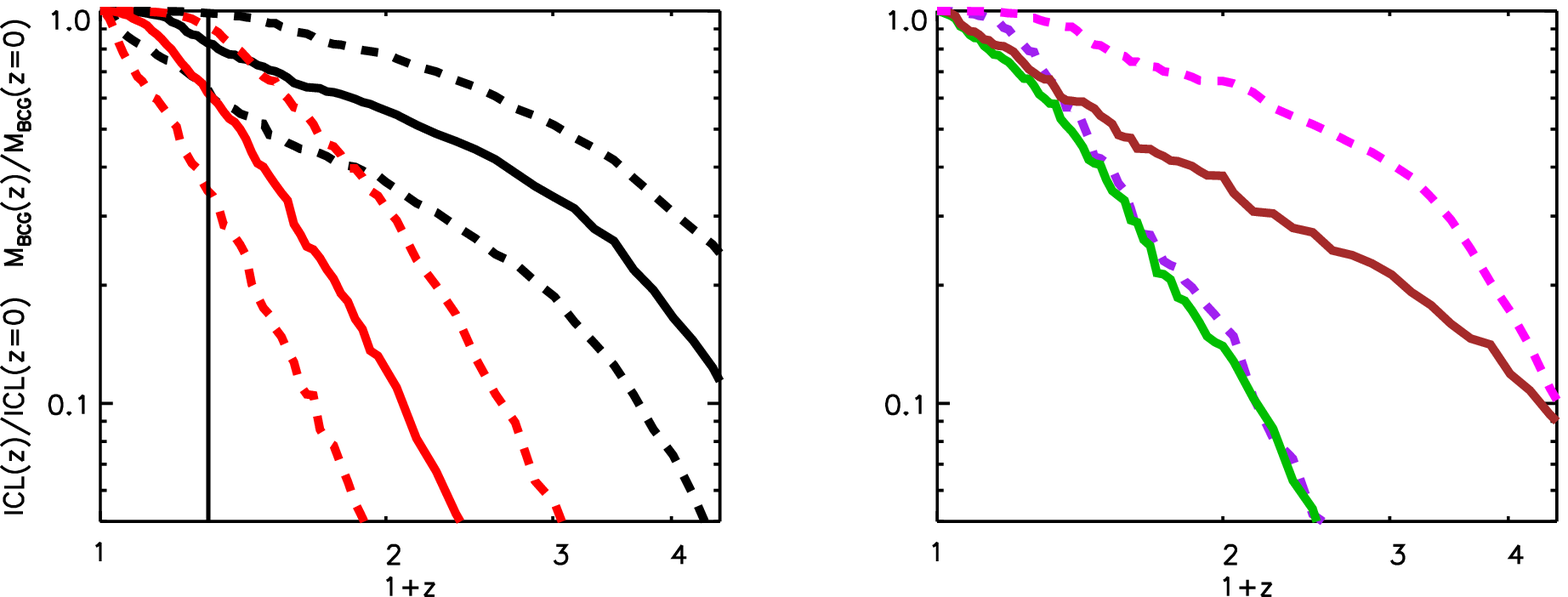} \\ 
\end{tabular}
\caption{ICL (red lines) and BCG (black lines) mass normalised to their value at $z=0$, as a function of redshift. The top panels show the prediction of the STANDARD models, for the whole 
sample of BCGs (left panel), and for two different bins in mass as indicated in the legend (right panel). Bottom panels: the same as the top panel, but for the MERGERS model. Solid lines 
represent the median of the distribution, and dashed lines represent the 16$^{th}$ and 84$^{th}$ percentiles.}
\label{fig:iclbcggrowth}
\end{center}
\end{figure*}

\begin{figure*} 
\begin{center}
\begin{tabular}{cc}
\includegraphics[scale=.8]{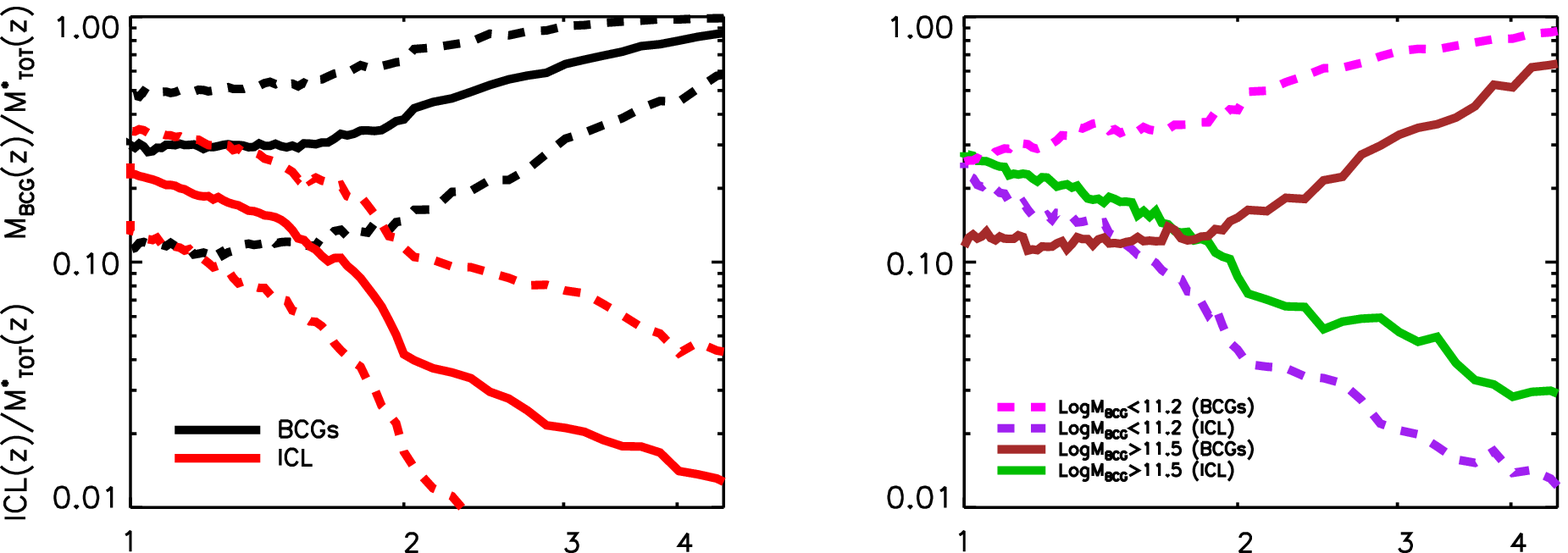} \\
\includegraphics[scale=.8]{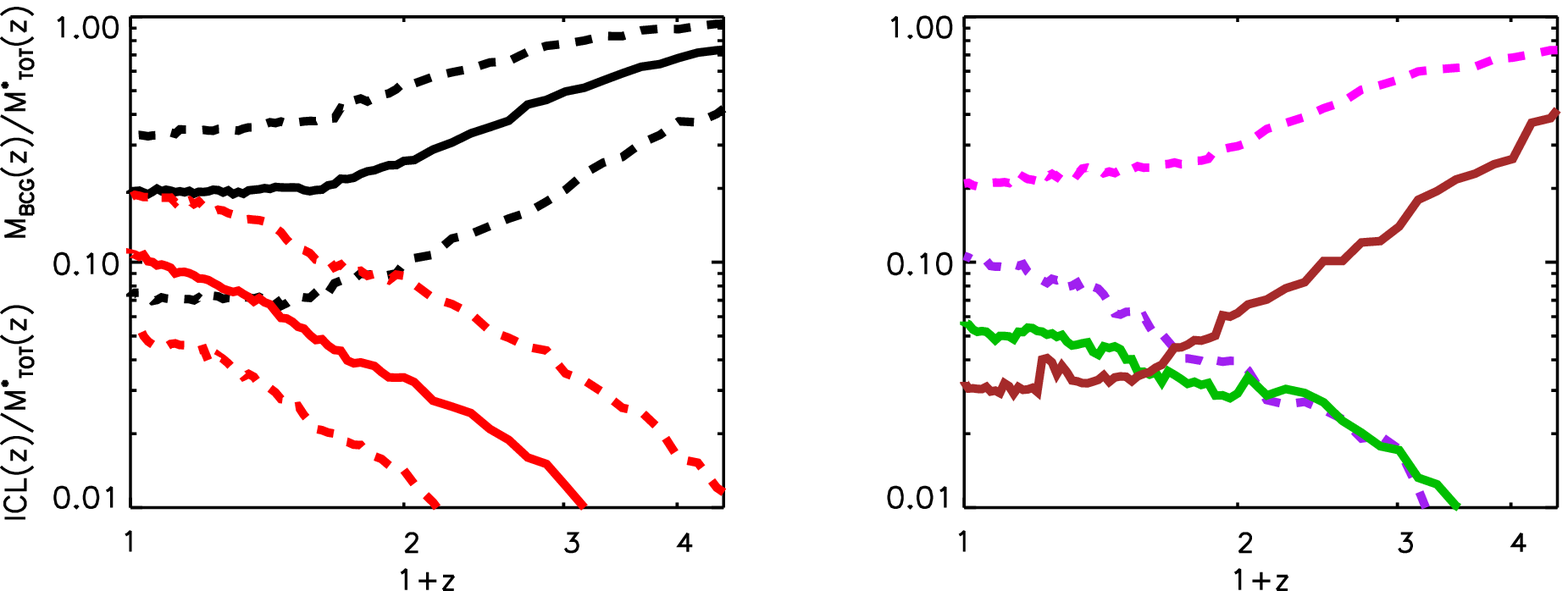} \\ 
\end{tabular}
\caption{Mass ratio between the ICL (red lines)-BCGs (black lines) at a given redshift and the total stellar mass within $R_{200}$ at the same redshift. The top panels show the prediction of 
the STANDARD models, for the whole sample of BCGs (left panel), and for two different bins in mass as indicated in the legend (right panel). Bottom panels: the same as the top panel, but for 
the MERGERS model. Solid lines represent the median of the distribution, and dashed lines represent the 16$^{th}$ and 84$^{th}$ percentiles.}
\label{fig:iclbcg_zeta}
\end{center}
\end{figure*}

\begin{figure*} 
\begin{center}
\begin{tabular}{cc}
\includegraphics[scale=.58]{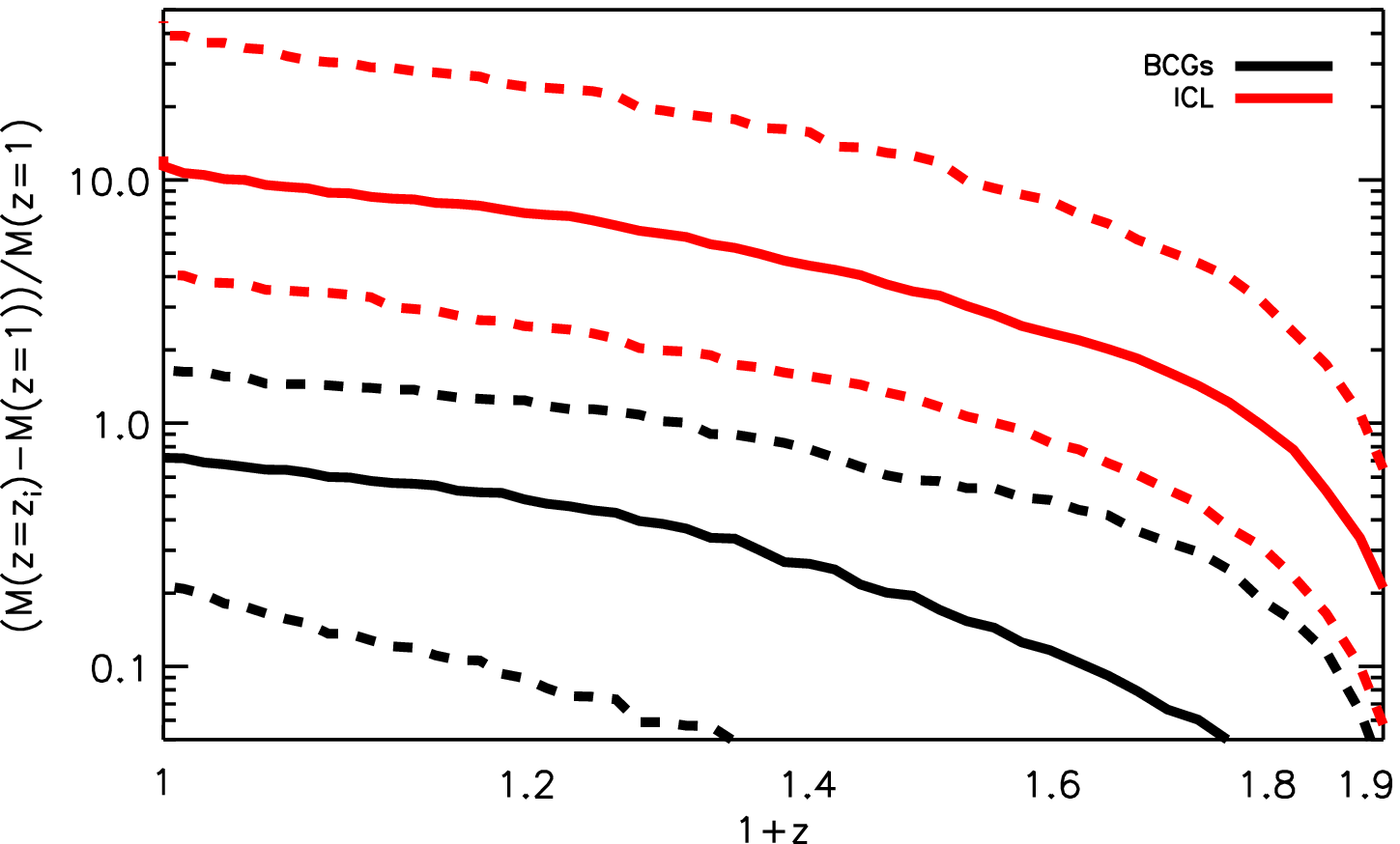} &
\includegraphics[scale=.58]{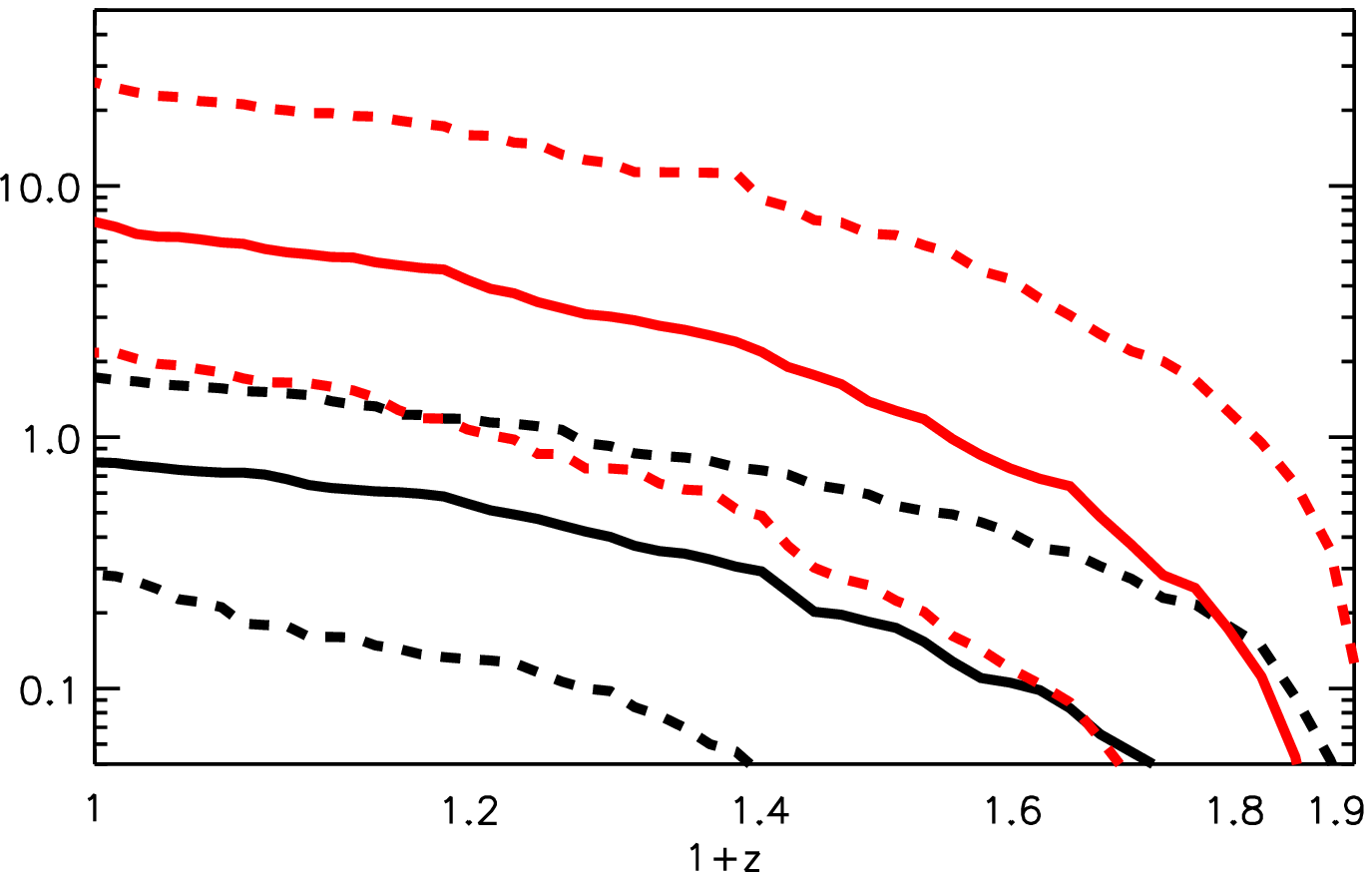} \\ 
\end{tabular}
\caption{Residual in mass ($z=z_i$)-($z=1$) of the ICL (red lines) and BCGs (black line) normalised to the mass at redshift $z=1$, predicted by the STANDARD model (left panel), and by the MERGERS 
model (right panel). Solid lines represent the median of the distribution, and dashed lines represent the 16$^{th}$ and 84$^{th}$ percentiles.}
\label{fig:iclbcggrowth_nz}
\end{center}
\end{figure*}

We now move our attention to the main goal of this study, that is, the ICL and BCG growth as predicted by the two models. In Figure \ref{fig:iclbcggrowth} we plot the mass of the ICL and BCGs 
as a function of redshift, normalised by their mass at the present time (formally identical to Figure 6 in C14) in the left panels, and the same quantities but for BCGs in the usual  
ranges of mass in the right panels. The top panels refer to the predictions of the {\small STANDARD} model, while the bottom panels refer to those of the {\small MERGERS} model. 
Let us focus on the differences between the top and bottom left panels. First, as already shown in C14, BCGs and their associated ICL evolve in different ways, and with different timescales. By 
the time BCGs have built-up around 60\% of their stellar mass, the ICL is just starting to assemble its, and this happens at around $z \sim 1$, for both models. Second, even though weak, there is 
a difference between the two models for the growth of the ICL. The vertical solid line separates two redshift ranges: from $z=0$ to $z=0.3$, and from $z=0.3$
to $z=3.5$ (the limit of the plot). The prediction of the {\small MERGERS} model for the growth of the ICL on the right side of the vertical line can be fit by a single power law. This means that the growth
of the ICL has no favourite timescale, in agreement with \cite{murante07}. However, in the case of the {\small STANDARD} model, a single power law appears not to be sufficient, which translates in 
a time dependent slope. This is an important message because, down to $z \sim 0.3$, the {\small MERGERS} model predicts a linear ICL growth in time, at odds to the 
{\small STANDARD} model which shows different ICL growth timescales. The trend is feeble, but visible.  

In the right panels we show the same quantities for BCGs and their ICL in the two mass ranges, $\log M_{BCG} < 11.2$ (magenta and purple lines) and $\log M_{BCG} > 11.5$ (brown and green 
lines). The trend discussed above is still visible, but there is no clear dependence on the BCG mass to which the ICL is associated. On the other hand, the BCG growth clearly depends on its 
final stellar mass. For both models, and even more accentuated in the {\small MERGERS} model, less massive BCGs grow much faster than more massive ones. At $z=1$, less massive BCGs already assembled 
around 70\% of their final mass, while more massive BCGs assembled just 40\%-50\%, depending on the model considered. 

We now want to quantify the growth of BCGs and ICL compared to the total stellar mass in the halo to which the BCG belongs, as a function of time. Such quantities, beyond the fact of being observable,
provide us a better understanding of the different growths of BCGs and ICL. In Figure \ref{fig:iclbcg_zeta}, we plot the ratio between the BCG and ICL stellar mass over the total stellar mass within the 
virial radius $R_{200}$, as a function of redshift. Panels and colors have the same meaning as in Figure \ref{fig:iclbcggrowth}. Both models predict a decreasing fraction for BCGs and an increasing fraction 
for the ICL. This is a natural consequence of the hierarchical growth of structures (see, e.g., \citealt{moster18}). The stellar mass within the virial radius keeps increasing with time, and it does faster than 
the BCG stellar mass. On the contrary, when the ICL starts its formation, it grows fast enough to be even faster than the global stellar mass growth within the virial radius. As done in the previous plots, 
we show also these quantities in different bins of the final BCG stellar mass. Again, less and more massive BCGs show different fractions, being lower for more massive BCGs. These central galaxies are supposed 
to reside in more massive haloes at any given time, and so the total stellar mass in these objects is higher. Interestingly, while for more massive BCGs the two fractions match at $z \sim 0.5$, this never happens 
for the less massive ones. We will discuss this in more details in the next section.

Clearly, BCGs and ICL, despite they have a common mechanism partially responsible for their growths, form and overall evolve differently, with different timescales. However, since the ICL starts to 
form much later with respect to BCGs, it is possible that, after a given redshift, the processes responsible for the growth of the two components lead them to a co-evolution. Let us consider the bulk of the 
ICL to form after $z=1$, as shown in C14 and in Figure \ref{fig:iclbcggrowth} of this work. In order to analyse the growth of BCGs and ICL not considering their evolution before $z=1$, and so to isolate 
the result of mergers (for both component), stellar stripping (for the ICL) and star formation (for BCGs) since $z=1$, in Figure \ref{fig:iclbcggrowth_nz} we plot the difference between the mass at any redshift 
and at $z=1$, normalised to the mass at $z=1$, for the ICL (red lines) and BCGs (black lines), as predicted by the {\small STANDARD} (left panel), and {\small MERGERS} (right panel) models. The predictions of 
both models show a steeper slope for the ICL in the very beginning of its formation, followed by a slow decline with time. After redshift $z \sim 0.7-0.8$, the slopes of the red solid line (ICL) and 
the black solid line (BCGs) are the same at any redshift. This is a clear evidence that, during the last 6-7 Gyr, BCGs and ICL co-evolve. We will come back on this in the next section.

\section{Discussion}
\label{sec:discussion}
As mentioned in Section \ref{sec:intro}, the scope of this work is to shed light on the BCGs and ICL growths by investigating on the relative contributions from the main formation channels of these 
components. The analysis done in Section \ref{sec:results} brought to a sequence of remarkable results, which we aim to discuss in more detail in the following. 

As yet, despite the copious amount of work done in studying the ICL, the scientific community has not reached a general consensus on the main process responsible for the ICL formation and growth. This 
turns out to be true, from observations to models. On the observational side, some authors invoke mergers as the dominant channel (e.g., \citealt{burke15,groenewald17}), others point on stellar 
stripping of satellite galaxies (e.g., \citealt{demaio15,montes18,demaio18,morishita17}), and all them by invoking similarities in quantities such as metallicity, age or colors between the ICL and 
stripped galaxies, or playing with the number of mergers and the contribution they can provide to the ICL, most of the times confiding in the help of theoretical models. 

By adopting dynamical arguments, \cite{burke15} find that, from $z \sim 0.9$ to $z \sim 0.1$, BCGs grow in stellar mass by a factor of 1.4 via mergers. The factor decreases to 1.2 if they assume that 
part of the mass of the satellite which merges with the BCG, specifically 50\%, goes to the ICL. With this assumption, which is common to most of the observational works, they find that the ICL grows 
in mass by a factor 4-5 in the same redshift range. Our {\small MERGERS} model, which makes use of the same assumption in the percentage of mass lost during each merger, predicts a similar growth for 
the ICL (a factor 6.1), and a factor 1.7 for BCGs, of which 92\% is given by mergers. Hence, our predictions are in good agreement with their findings, considering their limited sample and the scatter around 
our predictions. However, the growth factors predicted by the {\small STANDARD} model are very similar (6.6 for the ICL and 1.6 for BCGs, of which 88\% from mergers). It must be noted that in \cite{burke15} 
the authors consider only satellites within 50 Kpc from the BCG, with stellar mass ratios $M_{sat}/M_{BCG} <1$ and typical values for the circular velocity needed to compute the dynamical friction time. 
Their estimate gives a guideline of the growth factors, while our models consider all galaxies in a more consistent way. In fact, the authors conclude that the majority of the ICL must come from outside the 
core of the group/cluster, from mergers or stripping which contribute mostly to the ICL rather than the BCG.

Similar conclusions have been reached by \cite{groenewald17}. These authors focus on the importance of major mergers in the stellar mass build-up of a sample of BCGs between $0.1\lesssim z\lesssim 0.5$ 
and, by making the same assumption for the percentage of mass that goes to the ICL (50\%), conclude that major mergers contribute up to 30\% to the stellar mass of almost present day BCGs, since $z\sim 0.45$.
By the time, according to them, mergers bring sufficient stellar mass to the ICL to justify its growth down to the present day. They use our stellar mass growth published in C14 to infer the ICL mass growth.
Even though the method is inconsistent because the models presented in C14 considers both channels, their ICL growth factor agrees with ours and others in the literature. Moreover, it is also supported by the 
fact that both models presented here predict similar ICL growth factors, despite they are very different in spirit. 

The mergers channel has been ruled out by several authors, and for different reasons. Among all, arguments involving the metallicity, color and age of the ICL. These will be topics for further investigation 
and are, at the moment, beyond the scope of this paper. However, from the observationally side, there is evidence that the stellar stripping channel plays the dominant role. \cite{demaio15} analyse the BCG+ICL 
color gradient in four galaxy clusters at redshift $0.44<z<0.57$ and find, in three of them, a clear negative color gradient which decreases from 
super-solar in the BCG to sub-solar in the ICL. According to them, negative gradients can be the result of stellar stripping of $L*$ galaxies or disruption of dwarfs. The disruption of dwarfs can be discarded 
by invoking the constancy of the faint end slope of the luminosity function (\citealt{mancone12,stefanon13,demaio15}), and major mergers because they will tend to flatten the color gradient, and the number 
required to form the ICL is too high (\citealt{lidman13}). These results have been then confirmed by \cite{demaio18} and \cite{morishita17}, with similar arguments. \cite{montes18} study the ICL in six clusters 
at redshift $0.3<z<0.6$ and find that the average metallicity of the ICL is $\sim -0.5$, a result that along with the relative young stellar ages of the ICL they find, between 2 and 6 Gyr, is consistent with 
stripping of Milky Way-like galaxies accreted at $z<1$. All these results based on color gradients, metallicity and age of the ICL, are in good agreement with the predictions of the Tidal Radius+Merger model in
C14 that, we remind the reader, is identical to the {\small STANDARD} model in this work.

The {\small STANDARD} model is a combination of the two most relevant processes for the ICL formation. In Section \ref{sec:strip_contr}, we have shown that most of the ICL mass due to the stripping channel comes 
from the very innermost regions of the group/cluster, around 70\% within 100 Kpc from the centre. Most of this mass comes from stripping of intermediate/high mass galaxies (C14) through small/intermediate stripping events 
(Section \ref{sec:strip_contr}) and the main contribution is from disk-like galaxies, while ellipticals/spheroidals (B/T$>0.4$) contribute no more than 25\% (Section \ref{sec:strip_contr}). The prediction of our 
model concerning the main contributors to the ICL in terms of mass of the galaxy has then been confirmed by several observational works (e.g. \citealt{montes14,giallongo14,demaio15,annunziatella16,montes18,demaio18}), 
but also in apparent contrast with \cite{morishita17}, who find that low mass galaxies, $\log M_* <9.5$, contribute most to the ICL. No observational study so far has found any clue about the level of the events 
in terms of mass stripped during each event. This is reasonably arduous, if not impossible, just looking at the sky. However, from the typical color and metallicity of the galaxy which contribute most, it might 
be possible to discern between low-level stripping or high-level stripping events (low or high $M_{strip}^* /M_{gal}^*$ in the right panel of Figure \ref{fig:icl_BT}). \cite{demaio18} argue that the formation 
of the ICL depends on the types of galaxies, and given the fact that stripping is more 
efficient in the inner regions, as we confirmed, and the fact that in these regions early type galaxies are the dominant population, most of the ICL should come from this kind of galaxies. This is in contrast with 
what we stated above. Our model ({\small STANDARD}) predicts a larger contribution from disk-like galaxies close to the innermost regions. Despite they are less numerous, as explained in Section \ref{sec:strip_contr}, 
for morphological reasons they are more subject to stripping than ellipticals and spheroidals (see also \citealt{toledo11}). 

The central regions (the innermost 100-150 Kpc), are the place where most of the ICL form through tidal stripping. \cite{murante07}, taking advantage of hydrodynamic simulations and focusing on both channels for the ICL 
formation, find that 75\% of the ICL forms via mergers with the BCG or other massive galaxies, while stellar stripping has a minor effect. Their result is in contrast with the predictions of our {\small STANDARD} 
model. It must be noted that there is no conformity in the definitions, that is, our merger/stripping channels do not correspond to theirs. In fact, we consider as merger channel any ICL coming from satellites that 
merge with the BCG, and the merger happens in the very moment the dynamical friction time of the satellite goes to zero. In Murante et al., the merging process is a more continuous event with time and many 
particles that become unbound before the very moment of the merger (when the two galaxies are considered a unique object), are considered to belong to the ICL from the merger channel. This means, in a few words, 
that part of what they consider as merger channel is stripping in our model, and viceversa. We find that 70\% of the ICL from stellar stripping is produced in the innermost 100 Kpc, which is a fairly small region 
where a satellite galaxy can be "assumed" to be in the process of merging. In the {\small STANDARD} model, stellar stripping contributes to $\sim$85\% and mergers to $\sim$15\% to the total ICL. If we consider 
as "merger channel" the ICL that actually belongs to the stripping channel in the innermost 100 Kpc (in order to be more in line with their definitions), the percentages become 75\% for the merger channel and 25\% 
for the stellar stripping channel, in agreement with \cite{murante07}. Clearly, not all the ICL produced by stripping in the first 100 Kpc can be considered as "merger channel", because a given number of stripped 
satellites can just pass by the centre in radial orbits and survive for a long time. Hence, there still is an intrinsic inconsistency on the relative contributions of mergers and stellar stripping between our 
model and hydrodynamic simulations of Murante et al.. Moreover, it is worth reminding that our model considers all types of galaxies, while disk galaxies are not resolved in Murante et al.'s simulations. We have 
seen that disk-like galaxies play an important role in the stripping channel of our model.

A different modelling can also have consequences on the ICL properties that might be more qualitative rather than quantitative. In fact, Murante et al. find that the ICL has not a preferred timescale for its 
formation, which is also what our {\small MERGERS} model predicts. Nevertheless, despite the trend appears very weak, our {\small STANDARD} model predicts a rapid formation in the beginning that decelerates with 
time, implying that the ICL has a favorite timescale for its formation in the case stellar stripping is included.  

BCGs and ICL form, grow and overall evolve at different times and with different timescales. Nevertheless, at redshift $z \lesssim 0.7-0.8$, our models predict a clear co-evolution. The growth of BCGs is the one 
expected and predicted by the model of \cite{delucia07}, with different growth factors due to the implementention of ICL formation in our model. In Section \ref{sec:growth} we have discussed Figure \ref{fig:iclbcg_zeta}, 
which clearly shows the different growth pathways of ICL and BCGs. The fractions of stellar mass in the BCG/ICL with respect to the total stellar mass in the cluster evolve differently. The BCG accounts for the bulk of the 
stellar mass at high redshift, when the ICL has not started forming yet. The BCG contribution gradually decreases with time because more galaxies are accreted by the cluster, so the growth of the entire system is faster then 
the growth of the BCG itself. At lower redshift (depending on the model), the ICL starts to form, and its growth is very fast, faster than the growth of the whole system, making the contribution of the ICL to the total 
stellar mass being more and more important as time passes. 

However, as shown in Figure \ref{fig:iclbcggrowth_nz}, once the growth of both components before redshift $z \sim 1$ (which we consider as the formation time of the ICL, since most of it assembles later) is isolated, 
BCGs and ICL show a co-evolution down to the present time. During this time, BCGs grow mainly by mergers while the ICL assembles its stellar mass via mergers only in the {\small MERGERS} model, and also via stellar 
stripping in the {\small STANDARD} model. As discussed above, most of the ICL from stellar stripping comes from the innermost regions and a large part of it can be attributed to galaxies in the process of merging. It 
is, then, expected that the late growths of BCGs and ICL are connected, and linked to the dynamical history of the cluster.

The fractions of stellar mass in the BCG/ICL with respect to the total stellar mass is linked to the BCG mass at the present time, as shown by the right panels of Figure \ref{fig:iclbcg_zeta}. For the highest stellar mass 
BCGs, ICL and BCG mass fraction match at redshift $z \sim 0.5$ in the case of the {\small STANDARD} model, and somehow at lower redshift in the case of the {\small MERGERS} model, while in lowest stellar mass BCGs they are 
matching at the present time in the {\small STANDARD} model, but are still different in the {\small MERGERS} model.

Qualitatively speaking, this picture is in line with most of the observations (e.g., \citealt{burke15,morishita17}) and theoretical predictions (e.g., \citealt{tang18}) which focused on the ICL fraction in clusters 
as a function of time. \cite{burke15} find that ICL and BCGs contain similar fractions of the total light at $z \sim 0.4$. After this redshift, the ICL fraction grows very fast, while the BCG fraction stays constant, as in 
our models. Hence, as the authors point out, BCGs assemble before $z \sim 0.4$, and whatever happens after that redshift (they invoke mergers), is causing the growth of the ICL rather than the growth of the BCG. Similar 
results are found in \cite{morishita17}, where the ICL fraction at $z \sim 0.45$ is between 5\% and 20\%, and it is half of the observed fraction at the present time. The authors rightly conclude that this implies two 
distinct formation histories of BCGs and ICL. In another recent observation, \cite{montes18}, a slight increase of the ICL fraction with redshift is seen, but the authors do not confirm it because the redshift range 
explored in that work is too narrow to make strong conclusions. 

In order to definitively understand what process plays the most important role in the formation of the ICL, we need a clear observational evidence. Despite both {\small STANDARD} and {\small MERGERS} models predict 
different trends in many cases, the scatter around the median quantities are always large, and strong conclusions are not allowed. However, the ICL in groups and clusters is not only associated to the central galaxy, 
but it is found also around many other massive and less massive galaxies. We have shown (in Figure \ref{fig:iclbcg_icltot}) that the two models predict different trends and values for the ICL associated with the BCG 
over the total within the group/cluster. On cluster scale ($\log M_{200} \sim 15$), the two models predict rather different percentages, $\sim$80\% ({\small STANDARD}) and $\sim$45\% ({\small MERGERS}). Future deep
observations of local clusters can address this point and measure the contribution of the ICL associated with the BCG with respect to the total. Such a measurement can finally shed light on the relative importance of 
mergers and stellar stripping in the formation of the ICL.

\section{Conclusions}
\label{sec:conclusion}

We have studied the growth pathways of BCGs and ICL by means of a semi-analytic model of galaxy formation coupled with a set of high-resolution N-body simulations. In particular we have focused on the contribution 
of mergers and stellar stripping in building-up the ICL, and the relative contribution of mergers in the build-up of BCGs, addressing the point whether BCGs and ICL have similar formation histories or not. We have taken 
advantage of two different prescriptions for the ICL formation, one that considers both mergers and stellar stripping (named {\small STANDARD} model), and one that considers only mergers (named {\small MERGERS} model). 

According to our analysis and results, we conclude the following:
\begin{itemize}
 \item BCGs and ICL co-evolve during the last 6-7 Gyr, when the ICL is assembling the bulk of its mass. However, these two components form, grow and overall evolve at different times and with different timescales, 
         meaning that they have different growth pathways which depend on the dynamical history of the cluster.
 \item Mergers are an important process for the growth of both BCGs and ICL. On average, they account for half of the present day stellar mass of BCGs and the total (in the {\small MERGERS} model), or $\sim$15\% (in the 
       {\small STANDARD} model) of the stellar mass in the ICL. However, the fraction of stellar mass contributed by mergers depends on the present day mass of the BCG. More massive BCGs build-up a larger amount of mass
       via mergers than less massive ones. In the case of the {\small STANDARD} model, the percentages are 70\% and 35\% for more and less massive BCGs, respectively. Same trend and slight different percentages are found 
       in the case of the {\small MERGERS} model. On the contrary, the ICL build-up from mergers does not show any preferred stellar mass scale and thus it does not depend on the present day stellar mass of the BCG with 
       which it is associated. 
 \item Stellar stripping has the most important role for the formation of the ICL in the {\small STANDARD} model. Around 90\% of the ICL from stellar stripping is built-up in the innermost 150 Kpc. There is a BCG mass 
       dependence: more massive BCGs accumulate 75\% of their ICL, while the percentage is 94\% for the less massive ones. Most of the ICL coming from stellar stripping in the innermost 100 Kpc is from disk-like galaxies 
       (B/T$<$0.4), and ellipticals/spheroidals contribute no more than 25\%. Moreover, this ICL is the result of a large number of small/intermediate stripping events ($M_{strip}/M_{gal}<0.3$), while disruptive events 
       are very unlikely.
 \item The fraction of stellar mass in BCGs and in ICL over the total stellar mass within the virial radius evolve differently with time. At high redshift, the BCG accounts for the bulk of the mass, but its contribution 
       gradually decreases with time until it stays constant after $z\sim 0.4-0.5$. On the contrary, the ICL grow very fast and its contribution keeps increasing down to the present time. Again, these quantities depend on 
       the present day BCG stellar mass, in such a way that more massive BCGs contribute less to the total stellar mass at any time. For these centrals, the ICL and BCG fractions match at redshift $z\sim 0.5$. For less 
       massive BCGs, the two fractions match only at the present time and in the case of the {\small STANDARD} model.
 \item The ICL appears to grow with no favourite timescale in the case of the {\small MERGERS} model. Nevertheless, this is not the case in the {\small STANDARD} model. In fact, despite the trend is weak, the ICL growth 
       is faster in the beginning of its formation. 
 \item The two models, {\small STANDARD} and {\small MERGERS}, make predictions very similar in most of the cases, and often no strong conclusion can be taken when considering the scatter. However, the {\small MERGERS} model
       predicts a higher amount of ICL associated to other galaxies within the virial radius of the group/cluster other than the BCG on cluster scale, at $z=0$. For $\log M_{200} \sim 15$, the percentage of ICL in the BCG is 
       $\sim$80\% in the {\small STANDARD} model, and reduces to $\sim$45\% in the {\small MERGERS} model. We stress that this quantity can be constrained by future deep observations of local clusters, and such a 
       measurement can shed light on the relative importance of mergers and stellar stripping for the formation of the ICL.
\end{itemize}

In a forthcoming paper we aim to use these models, and variations of them, to study specific observable properties of the ICL. We will mainly focus on the color and metallicity of the ICL and how they compare 
with the measurements available in the literature.

\section*{Acknowledgements}
We thank the anonymous referee for comments and suggestions that improved the manuscript. 
S.K.Y. and E.C. acknowledge support from the Korean National Research Foundation 
(NRF-2017R1A2A1A05001116) and from the Brain Korea 21 Plus Program (21A20131500002). This study was 
performed under the umbrella of the joint collaboration between Yonsei University Observatory 
and the Korean Astronomy and Space Science Institute. X.K. acknowledges financial support by
the 973 Program (2015CB857003) and the NSFC (No.11333008).

\label{lastpage}

\bibliographystyle{mn2e}
\bibliography{biblio}

\end{document}